# Mutually enhancing light emission between plasmonic nanostructures and fluorescent emitters

Jingyi Zhao,[1,†] Yuqing Cheng,[1,†] Hongming Shen,[1,†] Yuen Yung Hui,[2] Huan-Cheng Chang,[2] Qihuang Gong,[1,3] and Guowei Lu[1,3,*]

[1]*State Key Laboratory for Mesoscopic Physics & Collaborative Innovation Center of Quantum Matter, Department of Physics, Peking University, Beijing 100871, China*
[2]*Institute of Atomic and Molecular Sciences, Academia Sinica, Taipei, 104 Taiwan, China*
[3]*Collaborative Innovation Center of Extreme Optics, Shanxi University, Taiyuan, Shanxi 030006, China*

We demonstrate that the fluorescent emitters can increase light emission from the plasmonic nanostructures in turn. With the help of atomic force microscopy, a hybrid system consisting of a fluorescent nanodiamond and a gold nanoparticle was assembled step-by-step for *in situ* optical measurements. We found that the emission from the nanoparticle increased compared with that before coupling. The interaction between plasmonic nanostructures and fluorescent emitter was understood as an entity based on the concept of a quantized optical cavity by considering the nanodiamond and the nanoparticle as a two-level energy system and a nanoresonator, respectively. The theoretical calculations reveal that both the plasmonic coupling effect and the dielectric nanoantenna effect contribute to the enhancement of light emission from the gold nanoparticles.

Enhancing light emission of fluorescent emitters with metallic nanostructures is key to a number of surface enhanced spectroscopies, such as surface-enhanced fluorescence (SEF).[1,2] Metallic nanostructures support localized surface plasmon (LSP) modes, which gives rise to a marked enhancement of local electromagnetic fields and results in a strongly modified local density of optical states.[3,4] By placing fluorescent emitters in the vicinity of the nanostructures, the light emission can be greatly enhanced because of the plasmonic resonant effect. It is well known that the plasmonic nanostructures enable the modification of the emission intensity, lifetime, directivity, and spectral shape.[5-11] Basic theories of the SEF process have been developed.[12-14] Plasmonic nanostructures known as optical antennas can be involved in both the excitation and emission process.[4,5,15-17] Regarding the interaction between the emitters and the plasmonic nanostructures, however, there is a lack of attention on how the emitters influence the light emission behavior of the plasmonic nanostructures in turn. Experimental access to another side of this process has long proven elusive.

Most of previous SEF theories often assumed that the light emission from plasmonic nanostructures is negligible and stable.[5,18-20] Such background emission is usually treated as a stable signal when calculating the SEF enhancement factor. Although this approximation allows us to understand the SEF main features well, a deviation of the enhancement factor is inevitable if the emission from the metal is omitted. Furthermore, this approximation would hinder us from understanding the SEF spectral shape fully and accurately.[4,21-23] Actually, intrinsic light emission from metal nanostructures has been demonstrated as a complementary property to absorption and scattering.[24-26] Although the mechanism is still in debate, a solid consensus is that metallic nanostructures can radiate at their LSP resonances after photon or electron excitation.[27,28] Then, light emission (here, photoluminescence, PL) from the nano-metal can be correlated with the broad "background continuum" in the SEF process.[29,30]

In this study, we demonstrate experimentally that the light emission from the plasmonic nanostructures is not a constant signal and it is changeable when coupling with the emitters. Which is to say that not only the metallic nanostructure enhances the light emission of the fluorescent emitters, but also the emitters increase the light emission from the plasmonic nanostructures in turn. With the help of an atomic force microscope (AFM), a single gold nanorod (GNR) was manipulated step-by-step to approach a single fluorescent nanodiamond (FND) on a glass coverslip. The light emission spectra excited with a CW laser were recorded *in situ* before and after coupling. We found that the light emission from the FND and GNR both increased after coupling. To understand the physical origin of this phenomenon, we give a theoretical description of the interaction between a two-level atom and a nanoresonator cavity based on the concept of a quantized optical cavity, as shown in Fig. 1. The model reveals that the plasmonic coupling effect can enhance light emission from the emitter greatly, but it also results in enhanced emission from the gold nanoparticles through LSP decay radiation. The induced local field of the nanodiamond felt by the gold nanoparticle also contributes to the enhanced emission from the nanoparticle. These findings reveal that the SEF phenomena is a mutually enhancing process between plasmonic nanostructures and fluorescent emitters, and suggest considering the plasmonic nanostructures and the emitters as a hybrid entity to analyze and optimize the surface enhanced spectroscopy.

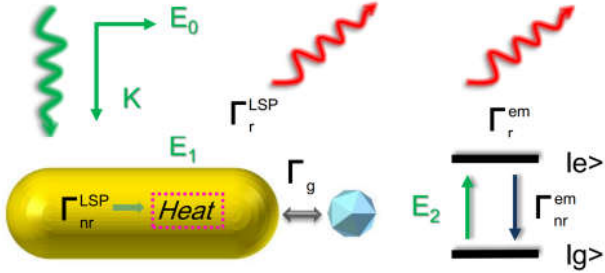

FIG. 1. (Color online) Scheme of the interaction between a FND and a GNR: the applied electromagnetic field $E$ induces polarizations that causes dipole-dipole coupling $\Gamma_g$, light emission $\Gamma_r^{LSP}$ from the GNR including both elastic and inelastic processes.

In our experiments, FNDs with a nominal size of 35 nm were produced by radiation damage of type-Ib diamond powders containing several nitrogen vacancy centers.[31] The GNRs were synthesized through a seed-mediated wet chemical method.[32] A dilute aqueous solution containing both the FNDs and GNRs was casted onto a silane-functionalized glass coverslip. Then, the nanoparticles were immobilized well, with an average spacing of several micrometers for single particle-level investigations. We characterized optically with a microspectroscopy system based on an inverted optical microscope combined with an AFM, as shown schematically in Fig. 2a. A CW laser at wavelength of 532 nm passing through an objective lens was used to excite the samples, and the fluorescent emission were recorded through the same objective lens. The light emission was recorded *in situ* before and after coupling. Moreover, the fluorescence signal from the same particles could be switched to an avalanche photodiode and analyzed with a TCSPC module (PicoHarp 300, PicoQuant). In this case, a picosecond laser diode operating at wavelength of 480 nm with repeat rate of 10 MHz was implemented for the fluorescence lifetime measurements. By AFM nano-manipulation, the GNRs were moved to approach the FNDs step-by-step, as shown in Fig. 2b. [33-37] The emission intensity and decay rate of the FNDs increased after being coupled with the GNR, which is indicated in Fig. 2d&e. In addition, the scattering spectra of the same particle can be obtained *in situ* by white light total internal reflection dark field method.

The interaction between fluorescent emitters and metallic nanostructures has been investigated extensively.[37,38] Despite much progress, even at the single nanoparticle level, the changes in plasmonic nanostructures' emission as influenced by the coupled fluorescent molecules were often ignored. This problem has not been addressed directly in the great number of recent studies, probably because of the experimental challenge of isolating the weak and complex signal from the SEF spectra. The PL from the metallic nanoparticles is usually too weak to recognize because the signal of the fluorescent emitters usually dominates the spectra in most SEF experiments. As a result, it is nontrivial to observe the intrinsic emission from the metallic nanoparticles modified by the coupled emitters. Here, the optical measurements were performed at the single particle level to avoid an inhomogeneous average from ensemble measurements. The PL spectrum of a GNR in Fig. 2d presents well-defined emission bands rather than a broad continuum band.[39] The emission spectra of the FNDs and GNRs are very stable (non-photobleaching and non-blinking),[40] as shown in Fig. 2c, which is important for *in situ* comparison before and after coupling.

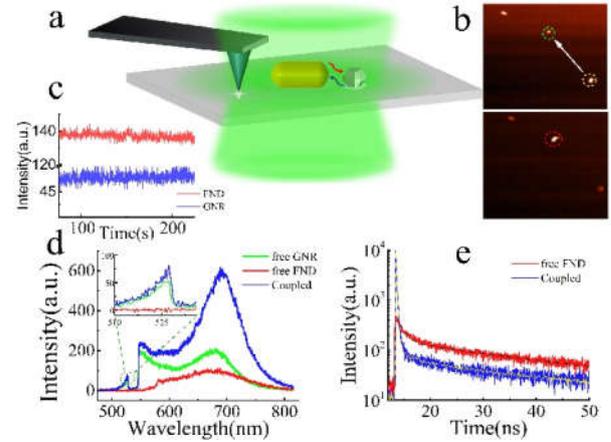

FIG. 2. (Color online) Scheme of nanomanipulation and representative in situ optical measurements. (a) Scheme of AFM manipulation method, (a) Representative AFM images during assembly process. d) Time trace of PL intensity of a GNR and a FND. (d) PL spectra for a free GNR (green) and a free FND (red) before coupling and that (blue) after coupling. Inset magnifies the anti-Stokes component. (d) Lifetime curves of the FND before (red) and after (blue) coupling with the GNR.

Specifically, as illustrated in Fig. 2d, the emission spectra of the FNDs dominate at the range of Stokes components but show no signal in the anti-Stokes range. In contrast, the emission spectra of the GNRs present an anti-Stokes band that decays exponentially as a function of photon energy. The one-photon luminescent anti-Stokes emission of plasmonic nanostructures has been studied in different systems, although the mechanism is still the subject of much debate.[27,29,30,39,41] Nevertheless, the anti-Stokes component allows us to recognize the intrinsic light emission of the GNRs from the SEF hybrid system. As shown in the inset of Fig. 2d, the intensity of anti-Stokes emission from the coupling system increases in comparison with that before coupling. The enhancement factor can be defined as $I_c/I_{uc}$, where $I_c$, $I_{uc}$ are the maximum of the PL spectra before and after coupling, separately. For the anti-Stokes emission, the intensity increased by approximately 40% at wavelength of 520 nm. The anti-Stokes emission is solely due to the GNR, which implies that the FND enhances the light emission from the GNR in turn. Regarding the Stokes, the spectrum of the coupled system undergoes a red shift, and the FND fluorescence intensity was enhanced as expected, even in comparison with the mathematic sum of the PL spectra of the free GNR and the free FND. Hence, the SEF phenomena is actually a mutual enhancing process due to the interaction between the fluorescent emitters and the plasmonic nanostructures.

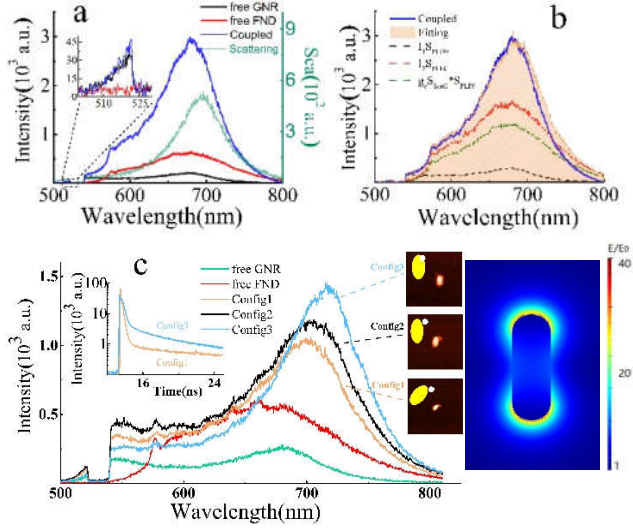

FIG. 3. (Color online) Understanding the SEF spectrum fully as three components decomposition. (a) PL spectra for a free GNR (black) and a free FND (red) before coupling, and the SEF spectrum (blue) and the scattering spectrum (green) after coupling. (b) Fitting spectra for $I_1 S_{PLG0}$, $I_2 S_{PLF0}$, and $g_c \hat{S}_{ScaG} \hat{S}_{PLF0}$, respectively. (c) Configuration dependent SEF spectra of a GNR and a FND coupling hybrid. Spectra for a free GNR (green) and a free FND (red) before coupling, and three SEF spectra for three different configurations as indicated by the AFM and schematic images. Inset shows lifetime curves of the config1 and config3 configurations for comparison, and near-field EM distribution around a GNR (40 nm×120 nm) at its longitudinal resonance.

To understand the SEF spectral shape fully, we correlated the PL and scattering spectra of the same single FND and GNR before and after coupling. We supposed that a SEF spectrum ($S_{SEF}$) contains three components: direct emission ($S_{PLG}, S_{PLF}$) from the GNR and FND, separately, and indirect emission ($S_{gc}$) through the coupling effect, i.e., $S_{SEF} = S_{PLG} + S_{PLF} + S_{gc}$. (It should be noted that, based on following quantum model, the direct emission ($S_{PLG}$) from the GNR can be understood as elastic radiation (i.e. plasmonic antenna effect) after coupling energy from local excitation field; the indirect emission ($S_{gc}$) is inelastic decay radiation of the GNR LSP resonance after coupling the excited states of the emitters.) And we assumed that the coupling rate ($\Gamma_g$) between the FND and the GNR is less than the internal phonon decay rate of the FND and the frequency dependent $\Gamma_g$ is strongly related with the LSP resonance, i.e., $\Gamma_g \sim \hat{S}_{ScaG}$ (the GNR scattering).[9,15,42] Then, we obtained the three components by fitting the SEF spectrum using $S_{SEF} = I_1 S_{PLG0} + I_2 S_{PLF0} + g_c \hat{S}_{ScaG} \hat{S}_{PLF0}$, where terms with 0 subscript are the spectra before coupling, $\hat{S}$ is the normalized spectrum, and $I_1, I_2$, and $g_c$ are constant parameters. Fig. 3a shows the experimental spectra $S_{PLG0}$ and $S_{PLF0}$ before coupling, and the $S_{SEF}$ which increased at both Stokes and anti-Stokes bands after coupling. The fitting results are summarized in Fig. 3b. The direct emission from the GNR and the FND increased by 40% and 140%, respectively, and the indirect emission photons caused by the antenna effect is approximately 30% of the whole SEF spectra. The emission from the metallic nanostructures can no longer be assumed to be a constant. And all three of these components determine the final SEF spectral shape for understanding the SEF spectral shape accurately.

In addition, we show that the SEF spectral shape, intensity, and lifetime vary for different coupling configuration.[15,43] It is well known that the coupling rate $\Gamma_g$ (i.e. Purcell factor) is dependent on separation and orientation of the dipole with respective to the nanoantenna. Then, the indirect emission ($S_{gc}$) is dependent on the configuration. The Purcell factor distribution is similar to the near-field distribution of a GNR as shown in Fig.3 inset.[42,44] Fig.3c shows the SEF spectra of another hybrid system with different configurations. As the FND is closer to the GNR end, the spectral maximum increases and red-shift more, and the lifetime is shorter. The FND as a dielectric particle would increase the local effective index felt by the GNR, resulting in a red shift of the LSP resonance, so do the SEF spectra. Moreover, the SEF of the FNDs coupling with gold sphere nanoparticles was also investigated in detail. Similar phenomena and conclusions were obtained as those by the GNR [Supplemental Material].

To understand the SEF mutual enhancement process, the hybrid of the GNR and FND was modeled as an entity based on the concept of a quantized optical cavity. We consider a FND interacting with a GNR separated by a distance (see Fig. 1). There is no direct electron tunneling between the GNR and the FND. The coupling mechanism is due to dipole-dipole interaction. The artificial hybrid system is excited with a quasi-continuous-wave laser beam with frequency $\omega_{ex}$ polarized along the system axis. Considering the GNR as a plasmonic resonator, the Hamiltonian of the resonator with LSP mode $a$ at $\omega_c$ is described as $H_c = \omega_c a^\dagger a$. Considering the FND as a two-energy-level atom system, the Hamiltonian is written as $H_m = \omega_g |g><g| + \omega_e |e><e|$, where the energy difference between $|g>$ and $|e>$ is $\omega_{em} = \omega_e - \omega_g$. Specifically, we define $\sigma_- = |g><e|$, $\sigma_+ = |e><g|$ as the transition operators. Therefore, the free Hamiltonian of the whole system without any interaction is written as: $H_0 = H_c + H_m = \omega_c a^\dagger a + \omega_g |g><g| + \omega_e |e><e|$. The interaction Hamiltonian is described as: $H_I = g(a^\dagger \sigma_- + a\sigma_+) + \mu_1 E_1(a^\dagger e^{-i\omega_{ex}t} + ae^{i\omega_{ex}t}) + \mu_2 E_2(\sigma_+ e^{-i\omega_{ex}t} + \sigma_- e^{i\omega_{ex}t})$, where the first term demonstrates that the LSP mode couples with states $|g>$ and $|e>$, and $g$ is the coupling constant. The second and the third terms show that the excitation electromagnetic field couples with the LSP mode and states $|g>$ and $|e>$, respectively, and $\mu_1$ and $\mu_2$ are the corresponding coupling constants. $E_1$ and $E_2$ are the respective localized electromagnetic field amplitudes that the GNR and the atom feel. Hence, the Hamiltonian of this system is given by $H = H_0 + H_I$. The dynamics of these modes can be solved by the equations $\dot{a} = i[H,a] - \kappa a = (-i\omega_c - \kappa)a - ig\sigma_- - i\mu_1 E_1 e^{-i\omega_{ex}t}$ and $\dot{\sigma}_- = i[H,\sigma_-] - \gamma\sigma_- = (-i\omega_{em} - \gamma)\sigma_- + iga\sigma_z + i\mu_2 E_2 \sigma_z e^{-i\omega_{ex}t}$, where $\kappa$ and $\gamma$ are the total decay of the GNR and the atom, respectively. $\sigma_z = |e><e| - |g><g|$, representing the difference between the level occupation numbers of states $|e>$ and $|g>$. Details of the solution are shown in the supplemental materials. We assume that $g \ll \kappa - \gamma$ because the SEF process is a weak coupling system.[43,45-47] This is supported by the fact that the frequency of the zero-phonon line of the FND does not shift after coupling with the GNR.

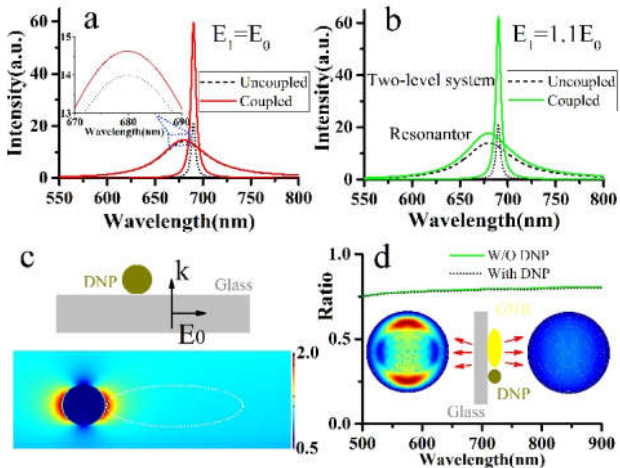

FIG. 4. (Color online) Theoretical understanding the influence of the coupling effect, the dielectric nanoantenna effect, and emission directivity on the GNR emission. Simulated spectra of a GNR (black dash) and an atom (black dot) for $g = 0$, and corresponding spectra (solid curves) after coupling for $g = 10\ meV$, (a) Local field felt by the GNR for $E_1 = E_0$, i.e., without the dielectric nanoantenna effect (b) for $E_1 = 1.1E_0$ due to the induced field of the FND. Inset in (A) presents an enlarged view of enhanced emission for the GNR. (c) Field distribution (X-Z plane) near a DNP with a diameter of 40 nm and refractive index of 2.3 by FDTD calculation, the averaged field felt by the GNR (white dot range) is higher than external excitation filed. (d) Ratio of emission flux from a GNR toward glass substrate to total flux with (black dot) or without (green solid) a DNP as function of wavelength. Inset in (D) shows representative emission patterns toward glass substrate or air as indicated at wavelength of 680 nm, respectively.

For $g = 0$, which suggests no coupling between the GNR and the FND, we obtain two Lorentz shape spectra as narrow emission from the atom and broad emission from the nanoresonator. When the coupling effect occurs ( e.g. $g = 10\ meV$ ) and the atom undergoes an enhanced excitation field because of the plasmonic near field of the GNR, then it is set as $E_2 = 1.5E_0$. We first assume that the field induced by the atom does not influence the local field felt by the GNR, i.e., $E_1 = E_0$. Then, the emission from the nanoresonator increases slightly (less than 10%), as shown in Fig. 4a. Then, the coupling effect cannot fully explain the total change in the experiment. Therefore, we increased $E_1$ and set it as $E_1 = 1.1E_0$. We obtain a considerable enhancement of the nanoresonator emission, as shown in Fig. 4b, which implies that the induced field produced by the polarization of the FND, i.e., a dielectric nanoparticle (DNP) as a non-plasmonic nanoantenna (i.e., dielectric nanoantenna),[48] could be the main factor for the enhanced emission from plasmonic nanostructure. This is supported by the FDTD numerical calculations shown in Fig. 5c. The DNP presents an enhanced local field that increases the EM field felt by the GNR. Hence, for $g \neq 0$, the light emission from the atom and nanoresonator both increase simultaneously. Both the antenna coupling effect and the dielectric nanoantenna effect contribute to this mutual enhancement. We noted that the influence of radiative directivity on the light collection efficient is faint. As shown in Fig.5d, the emission patterns of a GNR simulated with the FDTD method are almost the same before and after coupling with a DNP. The DNP decreases the EM flux toward the glass substrate only by less than 1% compared with that of a free GNR.

In summary, we have demonstrated in experiment and in theory that plasmonic nanostructures and fluorescent nanodiamond mutually enhance their light emission when they are coupled in the SEF process. This finding contributes to a full understanding of the SEF process. The SEF hybrid system should be considered as a whole entity for quantifying the SEF enhancement factor and their spectra shape. This observation introduces a new perspective in the field of surface enhanced spectroscopy. For instance, the mutual interaction is also effective in the surface enhanced Raman spectroscopy process.[49] The SEF background arising from the light emission of the metallic nanostructure is not a noise or constant background. The light emission from the plasmonic nanostructure is another indicator of the interaction strength between the plasmonic nanostructure and the emitter. And the SEF system should be considered as a hybrid entity to be analyzed and optimized.

## ACKNOWLEDGEMENTS

This work was supported by the National Key Basic Research Program of China (grant no. 2013CB328703) and the National Natural Science Foundation of China (grant nos. 61422502, 11374026, 61521004, 11527901).

[†]These authors contributed equally to this work.
[*] e-mail address: guowei.lu@pku.edu.cn